\documentclass[twocolumn]{aastex631}

\usepackage[T1]{fontenc}
\usepackage{graphicx}	
\usepackage{amsmath}	
\usepackage{amssymb}	
\usepackage{natbib}
\usepackage{multirow}
\usepackage{ulem}
\usepackage{bm}
\usepackage{stackengine}
\usepackage{xcolor}

\newcommand{\mbh}{$M_{\mathrm{BH}}$}

\newcommand{\Hb}{H{\small $\beta$}}
\newcommand{\Ha}{H{\small $\alpha$}}

\newcommand{\mmbh}{M_{\mathrm{BH}}}

\newcommand{\mMsun}{{\rm M_{\odot}}}

\newcommand{\ergs}{erg\,s$^{-1}$}
\newcommand{\kms}{km\,s$^{-1}$}
\newcommand{\mergs}{\mathrm{erg\,s}^{-1}}

\newcommand{\Qobj}{Q0957+561}

\shorttitle{Disk mapping of Q0957+561}
\shortauthors{Marculewicz et al.}

\begin{document}

\title{The disk reverberation mapping of the lensed quasar Q0957+561.}

\correspondingauthor{Mouyuan Sun}
\email{msun88@xmu.edu.cn}

\author[0000-0002-1380-1785]{Marcin Marculewicz}
\affiliation{Department of Astronomy, Xiamen University, Xiamen, Fujian 361005, People’s Republic of China; msun88@xmu.edu.cn}

\author[0000-0002-0771-2153]{Mouyuan Sun}
\affiliation{Department of Astronomy, Xiamen University, Xiamen, Fujian 361005, People’s Republic of China; msun88@xmu.edu.cn}

\author[0000-0002-2419-6875]{Zhixiang Zhang}
\affiliation{Department of Astronomy, Xiamen University, Xiamen, Fujian 361005, People’s Republic of China; msun88@xmu.edu.cn}

\author[0000-0002-5839-6744]{Tuan Yi}
\affiliation{Department of Astronomy, School of Physics, Peking University, Yiheyuan Rd. 5, Haidian District, Beijing, 100871, People's Republic of China}
\affiliation{Kavli Institute of Astronomy and Astrophysics, Peking University, Yiheyuan Rd. 5, Haidian District, Beijing, 100871, People's Republic of China}

\begin{abstract}
The measurement of continuum time lags in lensed quasars can effectively probe the accretion physics of quasars. This is because microlensing observations of lensed quasars can provide constraints on the half-light radii of quasar accretion disks. By combining the microlensing results with time lag measurements, one can, for the first time, estimate the propagation velocity of the physical process that drives inter-band time lags and cross-correlations among disk emission (i.e. in UV/optical bands). In this study, we perform the disk reverberation mapping study for the well-studied lensed quasar, \Qobj. The cross-correlation between the Zwicky Transient Facility (ZTF) $g$ and $r$ bands was measured; the $g$ variations lead the $r$ ones by $6.4\pm 2.6$ days in the rest frame. In combination with the half-light radius from the existing literature, we find that the propagation velocity of the variability mechanism should be $1.7^{+1.5}_{-0.7}$ times the speed of light. We discuss the possible outcomes of this result. Similar studies can be applied to other lensed quasars by utilizing the Legacy Survey of Space and Time (LSST) observations. 
\end{abstract}


\section{Introduction} \label{sec:intro}

Quasar accretion disks are too small to be spatially resolved; instead, their sizes can only be determined through two time-domain methods: microlensing observations and disk reverberation mapping. For gravitationally lensed quasars, microlensing due to stellar objects in the lensing galaxy, can induce extrinsic flux variations of the images of the lensed quasar. The flux variations can be used to measure the half-light radius of the supermassive black hole accretion disk \citep[e.g.,][]{Mortonson2005}. The measured half-light radii are three times larger than those predicted by the disk model, as evidenced by the findings of \cite[e.g.,][]{Morgan10, Fian21_15_fe}. In reverberation mapping (RM), the inter-band time lags between two or more continuum emissions are measured. It is speculated that a physical mechanism exists that drives disk emission variability from the inside out. Consequently, one would expect long-wavelength emission to lag behind short-wavelength emission with measurable time lags ($\tau$). If the physical mechanism responsible for the intrinsic UV/optical variability propagates across the accretion disk with a known speed, $v$, the relative distances between different emission regions can be expressed as $v\tau$. According to the X-ray reprocessing model (i.e., the illuminating variable X-ray emission is thermalized in the disk surface and reprocessed as a variable UV/optical emission; \citealp{Cackett2007reprocessing_model}), the disk RM relies upon the two implicit assumptions. First, the physical mechanism that drives the correlated variability in UV and optical wavelengths is the illumination of the accretion disk by X-ray emission \citep{Krolik1991, Cackett2007reprocessing_model}. Second, the propagation velocity ($v$) of the driving mechanism across the disk surface is the speed of light ($c$). As a result, one can use inter-band time lags ($\tau$) to infer disk sizes ($R = c \tau$). Previous studies \citep[e.g.,][]{Fausnaugh2016, Edelson2019, Guo2022, Fian2023_RM, MM_MS_23, Cackett2023, Sharp2024, Li2024} find that the measured time lags are also longer than the expectations from the classical disk model \citep[hereafter SSD;][]{SS73}. Here we propose that the propagation velocity of the driving mechanism across the disk can be inferred if the results from the disk RM and microlensing are combined. Hence, the disk RM studies of lensed quasars enable us to test the second aforementioned assumption (i.e., $v=c$) in the disk RM directly. 

Disk RM and microlensing observations are often performed in different bands. In order to combine the two results, it is necessary to use the temperature profile of accretion disk models to convert the measurements into the same wavelength. It is widely believed that luminous active galactic nuclei (AGNs) are powered by the SSD. According to the temperature profile of the SSD model, the emission-region sizes scale as \citep[e.g.,][]{Fausnaugh2016, Cackett2021_review}, 
\begin{equation}
    R = R_0 (\lambda/\lambda_0)^{4/3} \\,
\end{equation}
where $R$ and $R_0$ are the sizes for emission at wavelengths $\lambda$ and $\lambda_0$, respectively. One can use Eq. (1) to deal with the wavelength differences in the disk RM and microlensing observations. That is, one can calculate the corresponding light travel distance between two bands according to the half-light radius measurement in microlensing observations. Subsequently, the ratio of this light travel distance to the measured time lag in a disk RM is essentially the propagation velocity of the variability driving mechanism. We stress that this estimate of the propagation velocity is largely free of the highly uncertain black-hole mass or accretion rate. 

Several models have been proposed to explain microlensing or disk RM results. New models involving diffuse continua from inner broad-line regions \citep[BLRs; e.g.,][]{Korista2019}, disk structures significantly deviate from the SSD \citep{Hall2018, Sun2019winds, Starkey2023} or alternative driving mechanisms \citep{Gardner17, Cai2018_euclia, MSun2020_CHAR_model} with different $v$ are suggested. The propagation velocity $v$ of the driving mechanism is model-dependent. In some models, the X-ray emission is reprocessed by the SSD and the inner BLRs which produce diffuse continua \citep[e.g.,][]{Chelouche2019}, i.e., a significant fraction of the observed continua emission is not from the accretion disk. In this model, one still expects that $v=c$. In the EUV torus model of \cite{Gardner17} or the disk-corona magnetic coupling model of \cite{MSun2020_CHAR_model}, $v$ is significantly different from $c$. The disk mapping of lensed quasars enables us to determine the propagation velocity and test various models. 

\Qobj\ is a lensed quasar discovered by D. Walsh in 1979 with the cosmological redshift $z = 1.41$. This source has been the subject of extensive study with regard to lensing observations, \citep[e.g.,][]{Shalyapin2008, Hainline12_ml_event, Fian21_15_fe} and measurements \citep[e.g.,][]{Assef11, Fian22_mass}. According to literature, \Qobj\ harbors a supermassive black hole with several \mbh\ estimations, i.e., $\log (\mmbh/\mMsun)$=  $8.87 \pm 0.23$ and  $8.86 \pm 0.33$, for \Ha\ and \Hb\ by \cite{Assef11}, and $\log (\mmbh/\mMsun)$ $9.18 \pm 0.33$ by \citealp{Fian22_mass}. Furthermore, \citealp[]{Fian21_15_fe} determined the optical accretion disk of half-light radius $= 17.6 \pm 6.1$ lt-days.

In this work, we use the ZTF $g$ and $r$ light curves to determine their time lag for the lensed quasar \Qobj. This time-lag measurement and the half-light radius from microlensing observations \citep{Fian22_mass} is used to constrain the propagation velocity of the variability driving mechanism in \Qobj. Note that, \cite{Gil_td_2012} has obtained the $g$-$r$ ($4\pm1$ days) and $g$-$U$ ($3\pm1$ days) lags based on observations made with the Liverpool Robotic Telescope between December 26, 2009, and June 25, 2010. In comparison to the light curves presented in \cite{Gil_td_2012}, our ZTF light curves are considerably longer, allowing for the robust determination of time lags. Furthermore, \cite{Gil_td_2012} did not utilize the microlensing observations and inter-band lags to constrain the propagation velocity. 

This manuscript is formatted as follows. In Section \ref{sec:result}, we describe the methodology and data of the \Qobj, and present our results. In Section \ref{sec:discussion}, we discuss the physical implications of our results.

\section{Method and data} \label{sec:result}
\subsection{Photometetric data}
\begin{figure}
    \centering
    \plotone{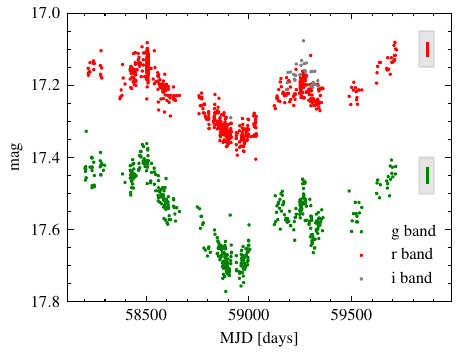}
    \caption{The \textit{g}, \textit{r}, and \textit{i}-band light curves of the Image 2 of the \Qobj\ from ZTF. Due to the limited coverage, we will not analyze the $i$-band light curve. The average error of each band is marked in grey boxes.}
    \label{fig:lc}
\end{figure}

Our light curves for \Qobj\ are based on observations obtained with the Samuel Oschin 48-inch Telescope and the 60-inch Telescope at the Palomar Observatory as part of the Zwicky Transient Facility (ZTF) project \citep{ZTF_data_processing2019}. ZTF monitors optical sources in $g$, $r$, and $i$-bands with coverage out to 30,000 square degrees of the Northern sky. We used ZTF DR19 data via the IRSA GATOR ZTF catalog service\footnote{\url{https://irsa.ipac.caltech.edu/cgi-bin/Gator}} to retrieve the ZTF $g$ and $r$ light curves of the two lensing images of \Qobj, i.e., Image 1 (R.A.: $150.3368$; decl.: $55.8971$) and Image 2 (R.A.: $150.3362$; decl.: $55.8988$). The duration of the light curves is $1895$ days in the observed frame. The average measurement error for the reported ZTF magnitude is $\delta m =0.02$ mag for the $g$ and $r$ bands. Hence, the signal-to-noise ratio for the observed flux is $\sim 50$ since $f/\delta f=1.08/\delta m$, where $f$ and $\delta f$ represent the observed flux and its measurement error, respectively. The $g$ and $r$ band light curves for image 2 are shown in Figure~\ref{fig:lc}. The $g$ and $r$ light curves exhibit prominent variability beyond measurement errors, which can be employed to constrain their time lag. The ZTF light curves of image 1 were not considered for the time-lag analysis due to the sparsity of the sampling. 

The ZTF light curves of \Qobj\ have several seasonal gaps. Consequently, the popular traditional linear interpolation adopted in PYCCF to fill the missing values in the unevenly sampled data and calculate the cross-correlation function can be inaccurate. In this work, we used the popular Damped Random Walk \citep[DRW; e.g.,][]{Kelly2009} model to describe and interpolate the ZTF light curves. First, we used \texttt{taufit}\footnote{https://github.com/burke86/taufit}, (see \citealt{Burke2021}) to find the best DRW parameters and their statistical distribution by applying the MCMC sampling. The DRW fitting code \texttt{taufit} is built upon the \texttt{celerite} \citep{celerite2017} that fits the model using Gaussian Process Regression. The logarithmic best-fitting damping timescale $\log (\tau_{\mathrm{DRW}}/\mathrm{days})$ are $2.95^{+0.59}_{-0.34}$ $3.17^{+0.57}_{-0.38}$ for $g$ and $r$ bands, respectively. The logarithmic best-fitting variability amplitude $\log \sigma_{\mathrm{DRW}}$ and their 1 $\sigma$ uncertainties are $0.42^{+0.15}_{-0.08}$ and $0.48^{+0.14}_{-0.09}$ for $g$ and $r$, respectively. Second, we adopted the best-fitting DRW parameters and the code \texttt{taufit} to generate evenly sampled $g$ and $r$ light curves. The sampling interval for the generated light curve is 0.3 days, and the duration of each generated light curve is equivalent to that of the actual data set. Third, for each generated light curve, we subtracted a second-order best-fitting polynomial function as often recommended to remove the long-term trend \citep[e.g.,][]{Welsh1999}. We stress that our results remain unchanged if no long-term trend is removed. Fourth, we obtain the cross-correlation function between the generated \textit{g} and \textit{r} light curves via PYCCF \citep{PyCCF_MSun_2018}. The time-lag ($\tau$) range measurement is from $-$100 to 100 days with a $0.3$-day step. In each run of the second step, the DRW models will generate different $g$ and $r$ light curves because the DRW is a stochastic model. To account for the uncertainties of the time lag due to the stochastic fluctuations, the whole process from the second to fourth steps is repeated $128$ times. The code to perform the aforementioned analysis can be downloaded from \url{https://github.com/marcinmarculewicz/Lensing_Quasar}. 

\begin{figure}
    \plotone{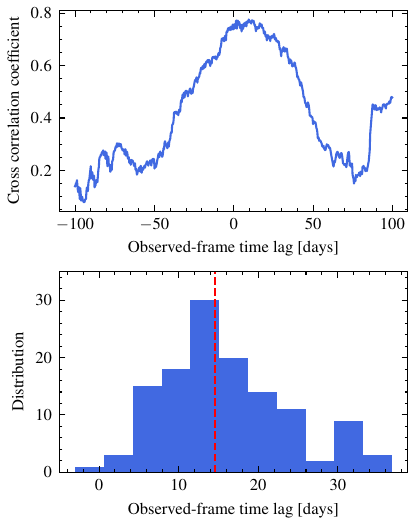}
    \caption{Top panel: The observed-frame time lag between $g$ and $r$ bands in days vs. correlation coefficient. Bottom panel: The $g$-vs-$r$ time lag centroid distribution. The red dashed line indicates the median value. It is evident that $g$ and $r$ light curves are correlated with a significant time lag of $15.4 \pm 6.2$ days.}
    \label{fig:time_lag_distrib}
\end{figure}

The advantage of the DRW approach is that this DRW model can fill well the gaps within the data set. Furthermore, we are also able to repeat the aforementioned five-step analysis many times to obtain the time-lay distribution and assess the time-lag uncertainty. The CCF function and time-lag distribution are shown in Figure~\ref{fig:time_lag_distrib}. For the $1\sigma$ uncertainty, we used 16 and 84 percentiles of the time-lag distributions. Hence, the measured time lag is $15.4\pm 6.2$ days in the observed frame. 

To check whether the DRW approach and linear interpolation approach are consistent, we performed the following analysis. First, we fit second-order polynomials to the $g$ and $r$ bands. Second, we subtract this second-order polynomial for each band. Then, the subtracted light curves are used to calculate the CCF via the linear interpolation. The linear interpolation and the DRW approach are in good agreement. Indeed, the observed time lag from the linear interpolation is $13.2\pm 3.4$ days, which is close to the result from the DRW method (i.e., $15.4\pm 6.2$ days).  

\begin{figure*}
    \plotone{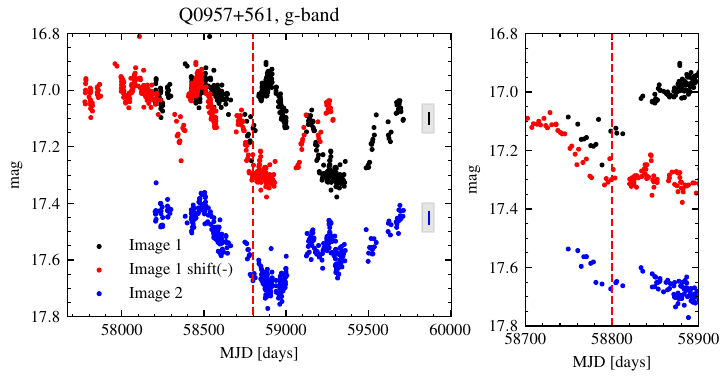}
    \caption{The ZTF $g$ band light curves of Image 1 (black dots) and Image 2 (blue dots). The possible microlensing event in \Qobj's ZTF $g$-band light curve. The average error of each band is marked in grey boxes. In both panels, the dashed red line illustrates the occurrence of the new microlensing event around MJD=58800 days.We shift the light curve of Image 1 (red dots) according to our measured time lag between the two images, which is $428 \pm 3$ days, see details in text. The right panel presents a zoomed-in view of the microlensing event around MJD=58800 days.}
    \label{fig:lc_ml}
\end{figure*}

There are reports of microlensing events in this source \citep[e.g.,][]{Hainline12_ml_event} and a subsequent analysis was conducted by \citealt{Fian21_15_fe}. The ZTF $g$ band light curve for Image 1 and Image 2 are shown in Figure~\ref{fig:lc_ml}. We find a new microlensing event in ZTF data. We measure the time lag between Image 1 and Image 2 of the ZTF $g$-band data via \texttt{PYCCF}. According to our \texttt{PYCCF} analysis, we obtained $428 \pm 3$ days. This value differs slightly from the previous estimation. For instance, \cite{Shalyapin2008} measured a time lag of $417 \pm 2$ days. After correcting for the time lag between the two images, we find and report the microlensing event starting from $\mathrm{MJD}\sim 58800$ (Figure~\ref{fig:lc_ml}). 

\subsection{P200 spectroscopic observations, fitting, and black hole mass calculation} 
We conducted spectroscopic observations of \Qobj\ using the P200 telescope on December 18, 2022. By rotating the slit orientation to align with the two lensed images of the object, the two images of \Qobj\ were simultaneously placed within the slit. Subsequently, we observed the object twice, with each exposed $1500$ seconds.

The one-dimensional spectra from the P200 observations were extracted using the IRAF v2.17 software \citep{iraf}. Standard procedures were employed for the extraction of the 1D spectra, including bias subtraction, flat correction, spectral extraction, wavelength calibration, and flux calibration. Given the proximity of the two images, which were separated by only 6 arcseconds, the same window was used for sky subtraction. However, different aperture positions were adopted to extract the counts spectra of the two images separately.

\begin{figure*}
    \centering
    \plottwo{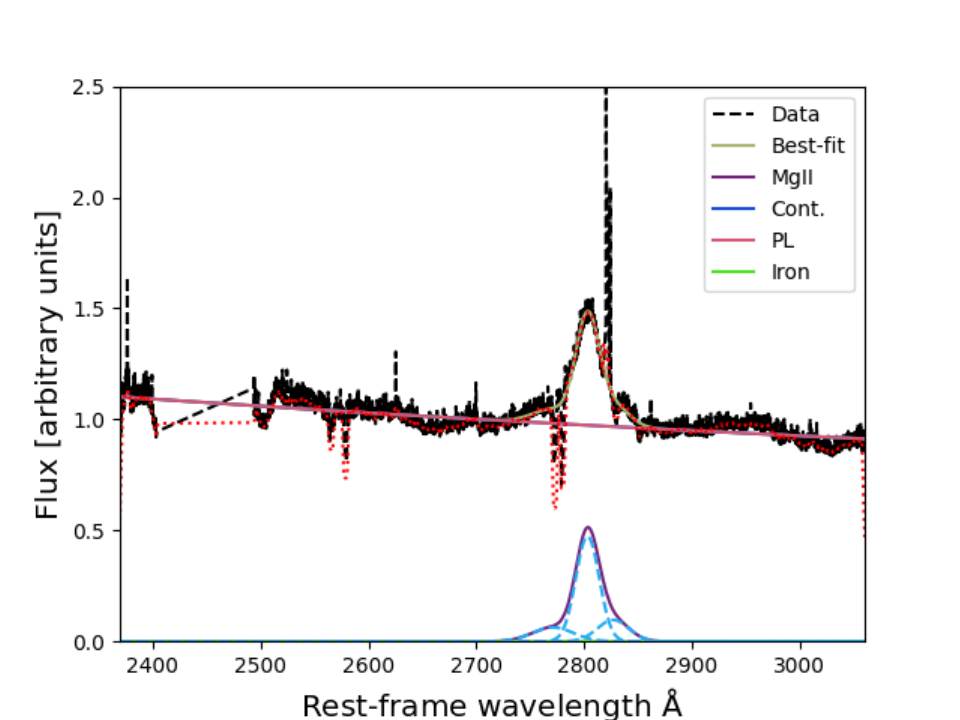}{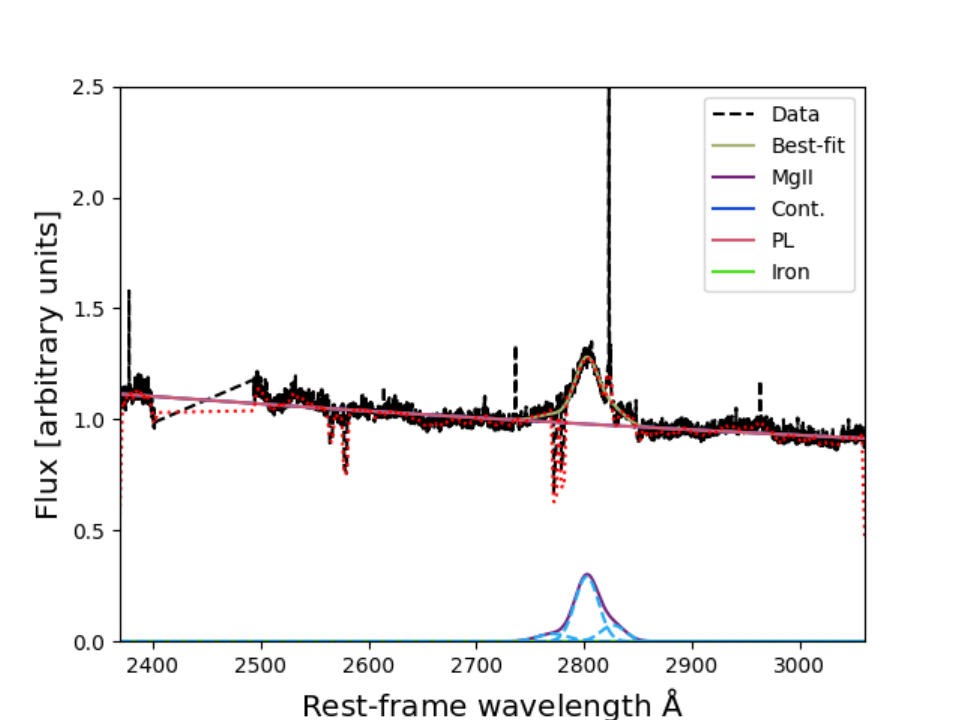}
    \caption{The spectroscopic decomposition and the best fits of the P200 spectra. The left and right panels are for Image 1 and Image 2, respectively. The black dashed and yellow solid curves represent the data and the best-fitting model, respectively. The blue and green curves represent the power-law continua and the optimal iron template, respectively. The purple solid curve represents the best-fitting \ion{Mg}{2} line profile that consists of three Gaussian components (blue dashed curves).}
    \label{fig:mg_fitl}
\end{figure*}

By performing P200 observations, we were able to obtain two new spectra of \Qobj\ (Figure~\ref{fig:mg_fitl}). Following \cite{Sun15}, we applied a pseudo-continuum model \cite[the power-law continuum with a Fe template from][]{VestergaardWilkes_fe} to fit the rest-frame line-free wavelength windows ($2250$--$2320$, $2333$--$2445$, $2470$--$2626$, $2675$--$2755$, $2855$--$3010$ \AA) by minimizing the $\chi^2$ statistic\footnote{The $\chi^2$ statistic is $ \chi^2 = \sum_{i=1}^n (O_i - E_i)^2/\sigma_{i}^2$ where $O_i$ is the $i$-th data point, $E_i$ is the model, $\sigma_i$ is the flux error, and $n$ is the total number of observed data of the spectra, respectively}. 

The \ion{Mg}{2} fit window is $2700$--$2900$ \AA. The best-fitting continuum model is used to calculate the underlying continuum within $2200$--$2900$ \AA\ range; then, the underlying continuum is subtracted from the spectrum to obtain the pure \ion{Mg}{2} line. In order to model \ion{Mg}{2} emission line, three Gaussian components were adopted. From the line fitting of \Qobj\ spectra and applying the minimizing $\chi^2$ statistic we obtained the Full Width at the Half Maximum (FWHM) of \ion{Mg}{2} = $3212\pm62$ \kms\ and $\lambda L_{\lambda} = (1.74 \pm 0.23) \times 10^{46}$ \ergs, $(2.20 \pm 0.23) \times 10^{46}$ \ergs, for Image 1 and Image 2 at 3000 \AA, respectively. Using our 3000 \AA\ luminosity measurement and the bolometric correction \citep[i.e., $5.62\pm1.14$;][]{Richards2006}, we obtain the bolometric luminosity $ L_\mathrm{{bol}}$. 

We collected historical literature on black hole mass estimations based on various emission lines. We included \Ha\ and \Hb\ black hole masses from \cite{Assef11} and \ion{C}{4} from \cite{Fian22_mass}. Using the recently updated and calibrated black hole mass from the Gemini Quasar Survey of \citeauthor[][(\citeyear{Dix2023}; see their Eq. (15))]{Dix2023},  we can calculate the \ion{Mg}{2} black-hole mass according to our \ion{Mg}{2} modeling results, i.e., $\log (\mmbh/\mMsun) = 9.13\pm 0.02$, $\log (\mmbh/\mMsun) =9.18\pm 0.02$ for Image 1 and Image 2, respectively. Our \ion{Mg}{2} masses are consistent with each other and are also statically consistent with the \Ha\ and \Hb\ masses of \cite{Assef11} and the \ion{C}{4} mass of \cite{Fian22_mass}. Across the whole paper we used the \ion{C}{4} black hole mass from \cite{Fian22_mass}. 

\begin{figure}
    \epsscale{1.2}
    \plotone{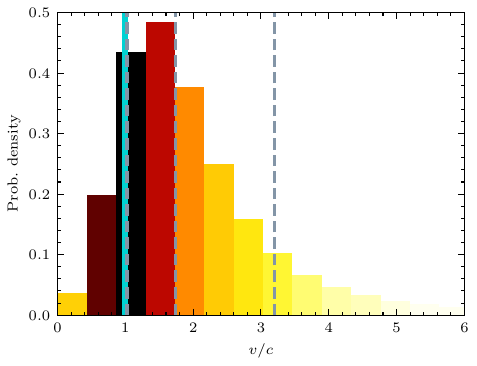}
    \caption{The probability density distribution of the inferred propagation velocity in units of the speed of light (i.e., $v/c$). Darker colors indicate higher probability densities. The three thin vertical gray lines represent the $16$-th, $50$-th, and $84$-th percentiles of the distribution. The thick cyan line indicates that $v/c=1$.}
    \label{fig:velocity}
\end{figure}

\section{Physical implications} \label{sec:discussion}

According to the X-ray reprocessing scenario \citep{Cackett2007reprocessing_model}, the theoretical time lag can be calculated for the SSD model \citep{SS73}. Using the black hole mass $\mmbh = 1.5\times10^9\ \mMsun$ \citep{Fian22_mass}, the luminosity at $5100$ \AA; $L_{5100} = 10^{45.79}\ \mergs$ \citep{Assef11}, the SSD model predicts that the characteristic radius ($R_{0,\mathrm{c}}$), where temperature satisfy the condition $kT (\lambda) = hc/\lambda$, is $R_{0,\mathrm{c}} = 5.8 \times10^{15}\ \mathrm{cm}$ at $\lambda_0=5100$ \AA\ (see Eq. (5) of \citealt{TK18}), where $k$, $h$, $c$, $T$, $\lambda$, are the Boltzmann constant, the Planck constant, the speed of light, the disk temperature, and the rest wavelength, respectively. In the X-ray reprocessing scenario, the CCF measures the variability weighted average emission-region light travel time, i.e., $R_0/c=5.04R_{0, \mathrm{c}}/c$ at $\lambda_0=5100$ \AA, where $c$ is the speed of light. Then, we use Eq. (1) to calculate the expected time lag between $g$ and $r$, which is $2.3$ days, hence the observed-frame time lag between $g$ and $r$ band is $5.5$ days. This value is statistically consistent within $2\sigma$ uncertainties with our measurement of $15.4\pm 6.2$ days. This finding is also consistent with the results in \cite{Li2021}, who found that the measured time lags are consistent with the SSD model for luminous quasars. 

The characteristic radius $R_{0,\mathrm{c}}$ can also be estimated from the microlensing observations. The estimation of $R_{0,\mathrm{c}}$ of the SSD model depends upon our knowledge of $M_{\mathrm{BH}}$ and $L_{\mathrm{bol}}$. It is widely known that the single-epoch mass estimators suffer from substantial uncertainties with the $1\sigma$ uncertainty to be $\sim 0.4$ dex \citep[e.g.,][]{Shen2011}. Hence, we aim to estimate $R_0$ from the half-light radius ($R_{\mathrm{f}}$) constrained via microlensing observations, i.e., $R_{\mathrm{f}}=2.44 R_{0,\mathrm{c}}$ \citep{Morgan10, TK18} if $T(R)\propto R^{-3/4}$. The half-light radius (i.e., the light produced within this radius is half of the whole disk emission at a given wavelength) measured by the microlensing observation of \Qobj\ is $R_{\mathrm{f}}=(4.6\pm 1.6) \times 10^{16}$ cm at $2558$ \AA\ \citep{Fian2023_Q0957}; hence, the corresponding half-light radius at $\lambda_0=5100$ \AA\ is $R_{\mathrm{f,0}}=R_{\mathrm{f}}(5100/2558)^{4/3}=(1.1\pm 0.4) \times 10^{17}$ cm. Then, $R_{0,c}=R_{\mathrm{f,0}}/2.44=(4.7\pm 1.6) \times 10^{16}$ cm according to the microlensing observation. The expected distance between $g$ and $r$ emission regions can be calculated from the microlensing observations. This is because $\Delta R_{\mathrm{var},gr}=R_{\mathrm{var},g}-R_{\mathrm{var},r}=5.04 R_{0,\mathrm{c}}((\lambda_r/\lambda_0)^{4/3}-(\lambda_g/\lambda_0)^{4/3})$ $=(2.9\pm 1.0)\times 10^{16}$ cm (see Eq. (1)). Again, we stress that this calculation is insensitive to our knowledge of $M_{\mathrm{BH}}$ and $L_{\mathrm{bol}}$. The calculation is always valid as long as the effective temperature profile of the disk scales as $T(R)\propto R^{-3/4}$. 

The relation between the expected distance $\Delta R_{\mathrm{var},gr}$ and the observed time lag $\tau_{g,r}$ is $\Delta R_{\mathrm{var},gr}=v\tau_{g,r}$. The propagation velocity we obtained is $v=\Delta R_{\mathrm{var},gr}/\tau_{g,r}=1.7^{+1.5}_{-0.7} c$ (Figure~\ref{fig:velocity}). Note that if we adopt the time lag from the linear interpolation, the estimated propagation velocity is $2.0^{+1.1}_{-0.7}c$, consistent with the DRW approach. 

In light of relatively large uncertainties, this result can be interpreted in several ways. The first possibility is that the apparent propagation velocity in fact, greater than the speed of light. Theoretically speaking, this result is possible and does not violet the special relativity as long as the X-ray reprocessing of a static SSD is revised. For example, \cite{Ren2024} point out that the X-ray reprocessing in the inhomogeneous accretion disk of \cite{DexterAgol2011} can result in large half-light radii and small continuum time lags. The presence of local temperature fluctuations in accretion disks results in a significant underestimation of the actual light travel time, as measured by inter-band time lags, for an inhomogeneous accretion disk \citep{Ren2024}. Hence, the apparent propagation velocity in this model would exceed the speed of light. An alternative hypothesis is that, the coupling between the corona and the disk is not driven by the X-ray reprocessing, but rather by magnetic coupling \citep[the CHAR model;][]{MSun2020_CHAR_model}. For luminous quasars, the inter-band time lags of the CHAR model can be very small, even smaller than the SSD predictions \citep{Li2021, Chen2024} because the emission regions of different bands are significantly overlapped. Hence, this model can also explain the calculated propagation velocity, especially if local temperature fluctuations are taken into account. The second possibility is that the derived propagation velocity is consistent with the speed of light. If this is the case, our results provide new evidence that the X-ray reprocessing is driving UV/optical variabilities in quasars \citep[but see][]{MM_MS_23}. 

We stress that the derived propagation velocity does not depend upon $M_{\mathrm{BH}}$ or $L_{\mathrm{bol}}$. Nevertheless, the result is sensitive to the temperature profile of the central engine and/or the origin of the continuum. For instance, additional contamination from BLRs may significantly affect the inter-band time lags, microlensing observations, and the estimations of the propagation velocity. Indeed, the response curves of the ZTF $g$ and $r$ bands cover CIII] and Mg II broad lines, respectively. Consequently, the measured half-light radius and inter-band time lag of \Qobj\ may not accurately probe the true disk sizes. \cite{Hainline12_ml_event} modeled the BLR contamination in the microlensing observations of \Qobj. Their result suggests that the estimation of the half-light radius is largely unaffected by the BLR contamination, provided that the latter contributes to $\leq 30\%$ of the observed flux. The BLR contamination might be important in the $g$-$r$ time lag \citep[e.g.,][]{Edelson2019, Guo2022, Korista2019}. Last but not least, it is also possible that the disk temperature profile is shallower than the classical $R^{-3/4}$ profile \citep[e.g.,][]{Sun2019winds, Li2019_micro_AD_wind, Cornachione2020_shallow_AD_temp, Zhou2024}. Nevertheless, the ratio of the half-light radius in a microlensing observation to the light-travel distance in a disk RM is largely insensitive to the temperature profile if the profile is a power-law \citep[ see e.g., Figure 1 of][]{Li2019_micro_AD_wind}. 

In the era of the Legacy Survey of Space and Time \citep[LSST;][]{LSST2019}, interband time lags and multiband microlensing half-light radii can be simultaneously measured for some lensed quasars. These results will enable us to reduce the majority of existing uncertainties and rigorously assess our conclusions.

\section*{Acknowledgments}

We would like to thank Jin-bo Fu for his help in obtaining the P200 data, and the anonymous referee for constructive comments that improved the manuscript. M.Y.S. acknowledges support from the National Natural Science Foundation of China (NSFC-12322303) and the Natural Science Foundation of Fujian Province of China (No.\ 2022J06002). 

Based on observations obtained with the Samuel Oschin Telescope 48-inch and the 60-inch Telescope at the Palomar Observatory as part of the Zwicky Transient Facility project \citep{ZTF-DOI}. ZTF is supported by the National Science Foundation under Grants No.\ AST-1440341 and AST-2034437 and a collaboration including current partners Caltech, IPAC, the Weizmann Institute for Science, the Oskar Klein Center at Stockholm University, the University of Maryland, Deutsches Elektronen-Synchrotron and Humboldt University, the TANGO Consortium of Taiwan, the University of Wisconsin at Milwaukee, Trinity College Dublin, Lawrence Livermore National Laboratories, IN2P3, University of Warwick, Ruhr University Bochum, Northwestern University and former partners the University of Washington, Los Alamos National Laboratories, and Lawrence Berkeley National Laboratories. Operations are conducted by COO, IPAC, and UW. Our P200 observation is kindly supported by the China Telescope Access Program (TAP).

\software{astropy \citep{astropy}, matplotlib \citep{matplotlib}, numpy \citep{numpy}, PYCCF \citep{PyCCF_MSun_2018}, emcee \citep{emcee}, IRAF \citep{iraf}}

\facilities{PO: 1.2m, Hale (Double Spectrograph)}

\bibliographystyle{aasjournal}
\bibliography{ref.bib}

\end{document}